\documentclass[letterpaper,onecolumn]{quantumarticle}
\pdfoutput=1
\usepackage{amsmath,amsthm,amsfonts,amssymb}
\usepackage{graphicx}

\usepackage{color}
\usepackage{hyperref}

\newtheorem{theorem}{Theorem}

\newtheorem{lemma}{Lemma}

\newtheorem{definition}{Definition}

\usepackage{algorithm}
\usepackage[capitalize,nameinlink]{cleveref}

\newcommand{\be}{\begin{equation}}
\newcommand{\ee}{\end{equation}}

\newcommand{\poly}{{\rm poly}}
\newcommand{\qpoly}{{\rm qpoly}}

\newcommand{\Pbias}{P_{coin}}
\newcommand{\Ocl}{{\rm AdNSP}}
\newcommand{\cH}{{\cal H}}
\newcommand{\isL}{{\cal L}}
\newcommand{\glen}{L}
\begin{document}

\title{The Power of Adiabatic Quantum Computation with No Sign Problem}

\author{Matthew B.~Hastings}

\affiliation{Station Q, Microsoft Research, Santa Barbara, CA 93106-6105, USA}
\affiliation{Microsoft Quantum and Microsoft Research, Redmond, WA 98052, USA}
\begin{abstract}
We show a superpolynomial oracle separation between the power of adiabatic quantum computation with no sign problem and the power of classical computation.
\end{abstract}
\maketitle

\section{Introduction}
The adiabatic algorithm\cite{farhi2001quantum} is a proposed quantum algorithm for optimization.  In its simplest form, one considers a quantum Hamiltonian which is a sum of two terms, one term being diagonal and proportional to the objective function of some classical optimization problem and the other so-called ``driving term" being a ``transverse field" (some non-commuting additional term).  One then adiabatically evolves the Hamiltonian from a large value of the transverse field (where the ground state is easy to prepare) to a small value of the transverse field, where the ground state encodes the desired solution of the optimization problem.

Unfortunately, there is tremendous theoretical evidence that gaps for random instances typically become super-exponentially small
\cite{altshuler2009adiabatic,altshuler2010anderson} so that the time required for adiabatic evolution to remain in the ground state is longer than the time required for even a classical brute force search\footnote{While some authors have disputed the perturbative calculation\cite{knysh2010relevance}, we believe that the general mechanism of localization on the hypercube will still apply for random instances with local driving terms.  The short path algorithm exploiting a difference between $\ell_1$ and $\ell_2$ localization may however be able give a super-Grover speedup\cite{Hastings_2018,Hastings_2019}.}.  Other authors have shown exponentially small gaps in simple problems\cite{laumann2012quantum} and some explicit simple examples show a super-exponentially small gap\cite{Wecker_2016}.  Further, separate from any question about the scaling of the gap, numerical experiments have shown that classical algorithms which simulate the quantum dynamics can perform comparably to a quantum device\cite{ronnow2014defining}.

Nevertheless, it remains of some interest to ask about the computational power of adiabatic quantum computation.  If we consider a Hamiltonian $H=sH_1+(1-s)H_0$ which is a linear combination of two {\it arbitrary} local Hamiltonians $H_0,H_1$, with some parameter $s$ controlling the dynamics, and require that the gap become only polynomially small so that the adiabatic evolution can be performed in polynomial time, then the problem is completely understood: this model is equivalent to standard quantum computation, and can solve any problem in BQP\cite{aharonov}.

However, if we restrict to the case that $H$ has no sign problem in the computational basis then the problem remains open.  Here ``no sign problem" means that in the given basis, all off-diagonal terms in $H$ are negative; this is sometimes termed ``stoquastic".
This case of no sign problem includes the adiabatic optimization algorithm discussed at the start if $H_1$ is equal to an objective function and $H_0$ is equal to $-\sum_i X_i$ with $X_i$ being the Pauli $X$ matrix on the $i$-th qubit.  It is important to emphasize that the arguments of \cite{altshuler2009adiabatic} apply even if the driving term has some sign problem; they depend rather on $H_1$ being an objective function and $H_0$ being chosen as some sum of local terms.

We remark that there is a problem of ``glued trees"\cite{Childs_2002} for which an exponential speedup over classical is known using a Hamiltonian with no sign problem that changes slowly in time\cite{Somma_2012}, but this problem is very different from the adiabatic annealing considered here.  The gap becomes exponentially small and the evolving quantum state has a large overlap with excited states during the evolution.  
More generally, if one allows dynamics in excited states, then it is possible to perform universal quantum computation using Hamiltonians with no sign problem\cite{Childs_2009}.  Thus, Hamiltonians with no sign problem are universal (using excited states) and adiabatic evolution is universal (using Hamiltonians with a sign problem).  The question we consider is what happens if we impose both these restrictions: no sign problem and adiabatic evolution in the ground state with a gap that is only polynomially small.

One piece of evidence that it may be hard to simulate this adiabatic evolution classically in general is the existence of topological obstructions to the equilibration of path integral Monte Carlo methods\cite{obs}.  As explained later, these obstructions help motivate the construction here.
Further, these obstructions are perhaps the main reason one should be interested in the question: topological obstructions such as difficulty equilibrating between different winding numbers can have an important effect on practical simulations of quantum systems as is well-studied in the condensed matter community\cite{Henelius_1998}, and so it would be useful if there were a general classical method that could overcome all such obstructions.

\subsection{Problem Statement and Results}
In this paper, we address this question.  Our results, in an oracle model defined below, will show a superpolynomial separation between the power of adiabatic computation with no sign problem and the power of classical computation.  At the same time, our results give no reason to believe that adiabatic computation with no sign problem is capable of universal quantum computation.

We use a number $N$ to parameterize the problem size (for example, in the example above using qubits, the number of basis states in the computational basis is $2^N$), and all references to polynomial scaling will refer to this parameter $N$.
We will assume more generally that the number of computational basis states is $\leq 2^{p(N)}$ for some function $p$ which is polynomially bounded,
 i.e., the computational basis states can be labelled using polynomial space.

We will define a path of Hamiltonians $H_s$, for $s$ in some interval to be {\it admissible} if it satisfies the following properties (when we refer to a parameter $s$ below, it is always assumed to be in the given interval).
First, for all $s$, $H_s$ must have no sign problem in the computational basis.
Second, for all $s$, $H_s$ must be polynomially sparse, meaning that for every computational basis element $|i\rangle$, there are at most
$\poly(N)$ basis elements $|j\rangle$ such that $\langle j | H_s | i \rangle$ is nonzero.
Third, for all $s$, for every $i,j$, $|\langle j | H_s | i \rangle| \leq \poly(N)$.
Fourth, for all $s$, $\Vert \partial_s H_s \Vert \leq \poly(N)$.
Fifth, for all $s$, $H_s$ has a unique ground state and the spectral gap to the first excited state is $\Omega(1/\poly(N))$.
Sixth, the length of the interval is $\poly(N)$.

Note that the number of basis states $2^{p(N)}$ and the definition of an admissible path both depend upon many polynomials that we have left unspecified.  The particular value of these polynomials is not important;
for example, for any such $p(\cdot)$ and $N$, we can define $N'=\lceil p(N) \rceil$, so that the number of basis states is $\leq 2^{N'}$.  Then, if the number of queries needed to solve all instances with some given probability is superpolynomial in $N$, it is also superpolynomial in $N'$.
Similarly, if that gap is lower bounded by $1/\poly(N)$ for some polynomial, it is lower bounded by $1/N''$ for $N''=\poly(N)$
and again the number of queries would be superpolynomial in $N''$.
Still these polynomials should be regarded as fixed in the main result: for some specific choice of polynomials, we show a superpolynomial number of queries.
For example, the construction we use gives
 $p(N)=O(N^2 \log(N)^3)$; it can be tightened somewhat.
 The reason it is convenient to leave these polynomials unspecified is that it simplifies some 
 accounting later:
 we will often, given some problem, construct a new problem with a larger number of basis states (increasing $p(N)$ so that it is still polynomial) or different gap; any change in the polynomials from this construction is often not stated explicitly.

The interest in the conditions other than the no sign problem condition is that adiabatic evolution on admissible paths can be efficiently simulated on a quantum computer up to polynomially small error, at least so long as the Hamiltonian is ``row computable", meaning that give any row index one can efficiently compute the nonzero entries in that row.  For us, we will consider an oracle problem where the oracle will give these nonzero entries.  Then, evolution under a time-dependent Hamiltonian for a time that is $\poly(N)$ will give the desired approximation to adiabatic evolution.

We will say that a path $H_s$ for $s$ in some interval $[a,b]$ satisfies the {\it endpoint condition} if 
for both $s=a$ and $s=b$, the ground state of $H_s$ is some computational basis state, i.e., for some $i$ (possibly different for $s=a$ and $s=b$), $|i\rangle$ is the ground state of $H_s$ (in a slight abuse of language, we will say that a vector is ``the" ground state of a Hamiltonian when of course the ground state is only defined up to phase).
We refer to $a$ as the start of the path and $b$ as the end of the path.
So, our interest will be in admissible paths of Hamiltonians which satisfy the endpoint condition because this gives a simple example of Hamiltonians for which it is easy to prepare the ground state of $H_a$ and for which one can measure in the computational basis\footnote{The reader might feel that the endpoint condition is a very strong restriction and wonder whether more general paths should be considered.  Of course, such more general paths may be useful in practice but since we are able to prove a superpolynomial separation even with this restriction, it seems worth keeping the restriction.  In particular, when the path satisfies the endpoint condition, it is very clear what it means to ``solve" the problem classically: one should be able to determine the computational basis state at the end of the  path as in \cref{docl}.}
 to determine the ground state of $H_b$.
We say that the path satisfies the condition at one endpoint if for one endpoint (either $s=a$ or $s=b$), the ground state is a computational basis state.

Also, we can easily concatenate admissible paths which satisfy the endpoint condition: given one such path $H_s$ for $s\in [a,b]$ with $|j\rangle$ being the ground state of $H_b$ and another admissible path $H'_s$ for $s\in [b,c]$ with $|j\rangle$ being the ground state of $H'_b$, we can concatenate the two paths to get a new admissible path if $H_b=H'_b$.  Even if $H_b$ differs from $H'_b$, it is possible to interpolate between $H_b$ and $H'_b$ by a path which first makes the off-diagonal elements of the Hamiltonian tend to zero, then changes the diagonal entries until they agree with diagonal entries of $H'_b$, then increases the off-diagonal elements until they agree with $H'_b$; doing this in an obvious way (for example, linear interpolation) still gives an admissible path from $H_a$ to $H'_c$.

One might be slightly surprised in one respect at our endpoint condition, though, since in adiabatic evolution with a transverse field, at $s=0$ the ground state of the Hamiltonian is a uniform superposition of all computational basis states, which may be written as $|+\rangle^{\otimes N}$.  However, for a system of $N$ qubits there is an obvious admissible path  $H_s=-(1-s) \sum_i Z_i - s \sum_i (X_i)$ for $s\in [0,1]$ with the ground state of $H_0$ being a computational basis state and the ground state of $H_1$ being $|+\rangle^{\otimes N}$.  So, given some admissible path which satisfies the endpoint condition at the end of the path and with the Hamiltonian at the start being $-\sum_i X_i$ (for example, interpolation between a transverse field Hamiltonian and some objective function for an optimization problem), we can concatenate with the path above to get an admissible path which satisfies the endpoint condition.

We will consider a version of the problem with an oracle in order to 
give a superpolynomial lower bound on the ability of classical algorithms to solve this problem.
Our oracle for a Hamiltonian will be similar to those considered previously\cite{Aharonov_2003,childs2004quantum,Berry_2006}.
The Hamiltonians that we consider can be simulated efficiently on a quantum computer with quantum queries of the oracle\cite{Aharonov_2003}.
As a remark for readers not familiar with oracle separations: to prove such a separation in a problem without an oracle (for example, when the Hamiltonian is a sum of two-body terms)
 would require proving that P is not equal to BQP, which is far beyond current techniques in computer science, although a separation between classical and quantum computation is known with respect to an oracle\cite{Simon}.
Also, an oracle separation suggests that if one wishes to have an efficient classical algorithm for the problem without an oracle, one should exploit some structure of the problem.

We define the oracle problem $\Ocl$ as follows\footnote{The terminology $\Ocl$ is an abbreviation of ``adiabatic no-sign problem".}.
\begin{definition}
\label{docl}
An instance of $\Ocl$ is defined by some admissible path
$H_s$ which satisfies the endpoint condition.
For definiteness, we assume that  $s$ lies in the interval $[0,1]$.
A query of the oracle consists of giving it any $i$ which labels some computational basis state $|i\rangle$ as well as giving it any $s\in [0,1]$, and the oracle will return the set of $j$ such that $\langle j | H_s | i \rangle$ is nonzero as well as returning the matrix elements $\langle j | H_s | i \rangle$ for those $j$ to precision $\exp(-\poly(N))$, i.e., returning the matrix elements to $\poly(N)$ bits accuracy for any desired polynomial.  We will call those $j$ the {\it neighbors} of $i$, and we will say that we {\it query state} $i$ at the given $s$.
The oracle will also return the diagonal matrix element $\langle i | H_s | i \rangle$.
The problem is: given query access to the oracle, and given the computational basis state $|i\rangle$ which is the ground state of $H_0$, determine the computational basis state $|j\rangle$ which is the ground state of $H_1$.
We say a classical algorithm solves this problem for an instance with some given probability if it returns the correct $j$ with at least that probability.
(As remarked above, the definition of an admissible path implicitly depends on various polynomials; so implicitly the problem $\Ocl$ also depends on various polynomials.)
\end{definition}

Note that given an unlimited number of queries to the oracle, it is possible to simulate the quantum evolution on a classical computer since one can determine the Hamiltonian to exponentially small error.

Remark: we have stated above that the oracle returns the matrix elements only to polynomially many bits.  This restriction is unnecessary for all the lower bounds on queries later, which would still hold even if the oracle returned the matrix elements to infinite precision.

Our main result is:
\begin{theorem}
\label{mainth}
For some constant $c$, for some specific choice of polynomials $p(\cdot)$ and choice of polynomials defining an admissible path, there is no algorithm that solves every instance of $\Ocl$ with probability greater than $\exp(-cN)$ using
fewer than $\exp(\Theta(\log(N)^2))$ classical queries.
\end{theorem}
Throughout, when we refer to an algorithm, the algorithm may be randomized and may take an arbitrary amount of time.
Remark: as is standard terminology, we refer to functions which are $O(\exp(\log(N)^\alpha)))$ for some fixed $\alpha$ as quasi-polynomial functions, and denote an arbitrary such function by $\qpoly(N)$.  A function which is 
$\exp(\Theta(\log(N)^2))$ is quasi-polynomial but is superpolynomial.

\subsection{Outline and Motivation for Proof}
The motivation for the proof of \cref{mainth} is, to some extent, an idea from \cite{obs}:  path integral Monte Carlo (which is a very natural classical algorithm for simulating quantum systems with no sign problem) in many cases cannot distinguish between the dynamics of a quantum particle on some graph $G$ and the dynamics on its universal cover $\tilde G$.  
The universal cover of a connected graph $G$ can be defined by picking an arbitrary vertex $v$ of $G$, and then vertices of the cover $\tilde G$ correspond to nonbacktracking paths starting at $v$, with an edge between vertices of $\tilde G$ if the corresponding paths on $G$ differ by adding exactly one step to the end of one of the paths.

However, at the same time, the largest eigenvalue of the adjacency matrix of $G$ (which will give us, up to a minus sign, the ground state energy of a Hamiltonian we define for that graph) may be much larger than it is on $\tilde G$ (to be precise, for an infinite graph we should not talk about ``the largest eigenvalue", but rather use spectral norm), so that a quantum algorithm can distinguish them.  We emphasize that if $\tilde G$ is a {\it finite cover} of $G$, the largest eigenvalue of the adjacency matrix of $\tilde G$ is the same as that of $G$.

This difference on infinite graphs has a finitary analogue: there is a difference between the ground state energy on a complete graph and the ground state energy on a tree graph with the same degree as the complete graph, with the difference in energy persisting no matter how deep the tree is, assuming the degree is $\geq 3$.  Here, the ground state energy of a graph is minus the largest eigenvalue of the adjacency matrix of that graph.

Our proof is based on the following main idea: we construct two different graphs which have different ground state energies, but for which we can give a superpolynomial lower bound on the number of classical queries to
distinguish those graphs; we quantify this ability to distinguish the graphs in terms of mutual information
between a random variable which is a random choice of the graphs and another random variable which is the query responses.  
We will term these graphs $C$ and $D$; these actually refer to families of graphs depending on some parameters.

The proof has two main parts: first, using these graphs to construct a family of instances of $\Ocl$ which cannot efficiently be solved with classical queries, and, second, proving the lower bound on the number of classical queries.  Proving spectral properties of the quantum Hamiltonians of these graphs is an additional part of the proof, but is relatively simple.

The first part of the proof is in
\cref{smqm}, where we show that it suffices to prove \cref{mainth} in a different query model.  
This part of the proof is perhaps less interesting than later parts of the proof, though it is important to understand the modified query model that we define.
In this query model, the oracle gives less information in response to queries, making it impossible to distinguish between a graph $G$ and some cover $\tilde G$.
Each vertex of the graph corresponds to a computational basis state so that neighbors of a vertex are also neighbors that one might receive in response to a query.  On $G$, one might follow some cycle on the graph, but if $\tilde G$ is the universal cover this is not possible: in the modified query model, one will not be able to know that has followed a cycle.

Then, in \cref{distg}, we then 
reduce the problem of proving \cref{mainth} to showing
two graphs $C,D$ satisfying certain properties exist.  The needed properties of the graphs are summarized briefly in \cref{trq}.  The main result is \cref{blemma}.

The rest of the paper is concerned with constructing $C,D$.
In  \cref{toomany} we give a first attempt at a construction, taking $C$ to be a complete graph of $O(1)$ vertices and $D$ to be a bounded depth tree graph (an infinite tree graph would give an example of a cover of $C$).
The idea is that if one does not reach a leaf of $D$, then $D$ is indistinguishable from the cover of $C$, and by choosing the
height of the tree superpolynomially large, it may take superpolynomially many queries to reach a leaf.

Unfortunately, the construction of \cref{toomany} suffers from two serious defects.  The first is that the it uses ``too many" states, i.e., number of vertices in the graph is not $\exp(\poly(N))$ if we take the height of the tree superpolynomially large.
The second and more serious defect is that the
graphs come with a privileged vertex called the ``start vertex", and in the modified query model we will still be able to determine when we return to the start vertex; a random walk in $C$ will often return to the start vertex but in $D$ one will not so a classical algorithm can efficiently distinguish them.
Nevertheless, this construction is worthwhile as it introduces certain key ideas used later.

The second part of the proof, constructing $C$ and $D$ fulfilling all needed properties (including that they cannot be efficiently distinguished by any classical algorithm and that there are only $\exp(\poly(N))$ vertices in the graph), starts in 
 \cref{decg}.  Here, we introduce the notion of a ``decorated graph".  The idea is to define some sequence of tree graphs for which it is ``hard" in some sense to reach certain vertices far from the start vertex because one tends (speaking very heuristically) to get ``lost" in other paths near the root.
 This tendency to get lost will make it hard for classical algorithms to detect the difference between $C$ and $D$.
 
In that section, we also give bounds on the energy of the Hamiltonians corresponding to these graphs and prove some properties of the ground states.

In  \cref{ch}, we give lower bounds on the number of classical queries needed to distinguish between $C$ and $D$.
\cref{distlemma} quantifies the difficulty of distinguishing $C$ and $D$ in terms of mutual information; \cref{distlemma} takes as input an assumption about difficulty of ``reaching" a certain set of vertices $\Delta$ in $D$ using queries starting from a given ``start vertex".  Difficulty of reaching this set follows from an inductive  \cref{inductivelemma}.

The results in  \cref{decg} and \cref{ch} are given in terms of a number of parameters.  In  \cref{chs} we fix values for these parameters and prove \cref{mainth}.

In \cref{linear} we briefly discuss a case of linear interpolation rather than arbitrary paths.

\section{Modifications to Query Model}
\label{smqm}
This section consists of two different subsections, which allow us to consider a more restrictive query model that we call the modified query model.

\subsection{Related States}
We first show that
we may, in everything that follows, assume that every state queried is either
 the initial state 
$|i\rangle$ which is the ground state of $H_0$ or a
neighbor of some state queried in a previous query.
For example, it may query $i$ for some value of $s$, receiving neighbors $j_1,j_2,\ldots,$.  It may then choose to query $j_1$ (possibly for some different $s$), receiving neighbors $k_1,k_2,\ldots$, at which point it may query any of $i, j_2,j_3,\ldots,k_1,k_2,\ldots$, but it will never query an ``unrelated state", meaning a state (other than $|i\rangle$) that it has not received in response to a previous query.

This result is essentially the same as Lemma 4 of the arXiv version of \cite{Childs_2002}.  The basic idea of the proof is one that we will re-use in \cref{mqm}.  At a high level, the idea is:
given any problem $H_s$ from an instance of $\Ocl$, we will construct some new oracle which is ``weaker" in some sense than the original oracle; in this case, whenever a related state is queried it returns the same responses as the original oracle, but whenever an unrelated state is queried, it returns some fixed response (i.e., the same response for any $H_s$) which hence gives no information about $H_s$.  Thus, queries of the weaker oracle can be simulated by queries of the original oracle (simply replace the response to a query of an unrelated state by this fixed response) but not vice versa.
We then construct (for any path $H_s$) some set of paths $H'_s$ which corresponds to some other instances of $\Ocl$ such that the original oracle for $H'_s$ is almost equivalent to the weaker oracle for $H_s$.  Here, ``almost equivalent" means that for a random choice of path $H'_s$ from this set, with probability close to $1$ the original oracle for $H'_s$ returns the same responses as does the weaker oracle for $H_s$.
Finally, if there is some algorithm $A$ that solves every instance of $\Ocl$ with probability $p$, including in particular the paths $H'_s$,
we can define an algorithm $A'$ which is given by algorithm $A$ using queries of the weaker oracle for $H_s$.  Then algorithm $A'$ will solve every instance of $\Ocl$ with probability close to $p$ using only the weaker oracle.
This construction will imply some change in the polynomials defining $\Ocl$; in particular, in this case the size of the Hilbert space will change.

Formally:
\begin{lemma}
\label{relstatelemma}
For any algorithm $A$ that solves every instance of $\Ocl$ with probability $\geq p$ using only quasi-polynomially many queries and possibly using queries of unrelated states,
there is some algorithm $A'$ which only queries related states and solves every instance of $\Ocl$ with probability $\geq p-\qpoly(N) 2^{-N}$ using at most as many queries as $A$ (albeit with some change in the polynomials defining $\Ocl$).
Hence, if algorithm $A$ succeeds with probability large compared to $\qpoly(N)2^{-N}$, then algorithm $A'$ succeeds with a probability that is comparable to that of $A$.
\begin{proof}
Given any path $H_s$ with Hilbert space $\cH$ of dimension ${\rm dim}(\cH)=2^{p(N)}$, consider a new Hilbert space $\cH'$ of dimension ${\rm dim}(\cH')=2^{p(N)+N}$ which is exponentially larger.  Define
a path $H'_s$ by
\be
\label{perminto}
H'_s=\Pi \begin{pmatrix} H_s & \\ & W I\end{pmatrix} \Pi^{-1},
\ee
where the rows and columns correspond to computational basis states, 
where the first block is of size $2^{p(N)}$ and the second block is of size $2^{p(N)+N}-2^{p(N)}$, 
where $\Pi$ is a permutation matrix chosen uniformly at random,
and where $I$ is the identity matrix
and $W$ is a scalar.  We choose $W$ larger than the largest eigenvalue of $H$ so that
the ground state of $\begin{pmatrix} H_s & \\ & W I\end{pmatrix}$ is completely supported in the first block and is given in the obvious way from the ground state of $H_s$, and hence the ground state of $H'_s$ is given by applying $\Pi$ to that state.

Suppose on the $q$-th query, the algorithm queries an unrelated state.  The number of unrelated states whose image, under $\Pi^{-1}$, is in the second block is at least ${\rm dim}(\cH')-{\rm dim}(\cH)-q$.
Since the permutation $\Pi$ is random, with probability at least $({\rm dim}(\cH')-{\rm dim}(\cH)-q)/{\rm dim}(\cH')$ the response to the query will be that the state has no neighbors and that the diagonal matrix element of that state is $W$.
Call this ``response $R$".

If an algorithm makes only quasi-polynomially many queries, with probability $\geq 1-\qpoly(N)2^{-N}$ the response to
{\it all} queries of unrelated states will be $R$.
So, given some algorithm $A$ which may query unrelated states, and which makes only quasi-polynomially many queries, we may define a new algorithm $A'$ which modifies $A$ by assuming (without querying the oracle) that the response to any query of an unrelated state will be $R$.  Remark: if $A'$ finds some inconsistency in this assumption, for example if it queries some unrelated state and assumes the response is $R$ and then later that state is returned as the neighbor of some previous query, algorithm $A'$ will terminate and return some arbitrary result.

Then, $A'$ queries only related states and, with probability $\geq 1-{\rm qpoly}(N) 2^{-N}$, algorithm $A$ returns the same result as algorithm $A'$ does.  Here ``probability" refers to both random choice of $\Pi$ and randomness in $A$; if $A$ is randomized, of course we assume that $A'$ uses the same source of randomness.
Each instance of $H'_s$ is defined by an instance of $H_s$ and by a choice of $\Pi$. 
If $A$ returns the correct result for every instance of $H'_s$ with at least probability $p$ for some $p$, then, trivially, for any $H_s$ the average over $\Pi$ of its probability of returning the correct result is at least $p$.  Hence, for any $H_s$, the probability that $A'$ returns the correct result is at least $p-{\rm qpoly}(N) 2^{-N}$.
\end{proof}
\end{lemma}

\subsection{Modified Query Model}
\label{mqm}
We now introduce the modified query model
in contrast to the query model given previously (which we will refer to as the original query model).
We show that if  \cref{mainth} holds using the modified query model, then it holds using the original query model.
Very briefly: the modified query model will be such that if the algorithm follows some nonbacktracking path of queries that forms a cycle (for example, querying $i$ to get some neighbor $j$, querying $j$ to get some neighbor $k$, querying $k$ to get $i$ which is a neighbor of $k$), then the query responses will make it impossible to determine that one has returned to the start of the cycle (in this case, $i$).

To explain the modified query model in more detail, we have an infinite set of {\it labels}.  Each label will correspond to some computational basis state, but the correspondence is many-to-one; we describe this correspondence by some function $F(\cdot)$.  The algorithm will initially be given some label $l$ that corresponds to the computational basis state $|i\rangle$ that is the ground state of $H_0$.  
A query of the oracle consists of giving it any label $m$ that is either $l$ or is a label that the algorithm has received in response to some previous query, as well as giving it any $s\in [0,1]$, and the oracle will return 
some set $S$ of labels such that $F(S)$ is
the set of $j$ such that $\langle j | H_s | F(m) \rangle$ is nonzero.  
Distinct labels in $S$ will have different images under $F(\cdot)$ so that $|S|$ is equal to the number of neighbors.

The oracle will also return, for each label $n\in S$,
the matrix elements $\langle F(n) | H_s | F(m) \rangle$ to precision $\exp(-\poly(N))$, i.e., returning the matrix elements to $\poly(N)$ bits accuracy for any desired polynomial.  
The oracle will also return the diagonal matrix element $\langle F(m) | H_s | F(m) \rangle$.
The labels in $S$ will be chosen as follows: if label $m$ was received in response to some previous query on a label $n$, so that 
$\langle F(n) | H_s | F(m) \rangle$ is nonzero and hence $F(n)\in F(S)$, then label $n$ will be in $S$, i.e., we will ``continue to label $F(n)$ by label $n$", or equivalently ``on a backtracking path, one realizes that one has backtracked".  However, for {\it all} other $j$ such that
$\langle j | H_s | F(m) \rangle$ is nonzero, we will choose a new label (distinct from all previous labels) to label the given vertex $j$, i.e., a new label $o$ such that $F(o)=j$.

Thus, after a sequence of queries by the algorithm, we can describe the queries by a tree, each vertex of which is some label, with neighboring vertices in the tree corresponding to computational basis states which are neighbors.

We use the same idea as in the proof of \cref{relstatelemma}.  In this case, the weaker oracle is the oracle of the modified query model.  This can clearly be simulated by the original oracle, since one can simply invent new labels for a state if the oracle gives one a label that one has seen previously, but not necessarily vice versa.

We define, for each Hamiltonian $H$,
 a model which has a large but finite set of labels.  The function
$F(\cdot)$ mapping labels to vertices will be $2^N$-to-one.  If Hamiltonian $H$ acts on Hilbert space $\cH$, then these labels $l$ will correspond one-to-one to computational basis states of some Hilbert space $\cH'$ with dimension $2^N {\rm dim}(\cH)$.
We will define a Hamiltonian $H'$ acting on $\cH$ as follows.

Label computational basis states of $\cH'$ by a pair $i,x$ where $i$ is a computational basis state of $\cH$ and $x$ is a bit
string of length $N$.
For each $a\in \{0,1,\ldots,N-1\}$, let $X_{a}$ denote the operator that flips the $a$-th
bit of this bit string, i.e.,
$$
X_a=\sum_{i,x} |i,x\oplus 1_a\rangle \langle i,x|,$$
where $1_a$ is a binary vector with entry $1$ in the $a$-th position, and $0$ elsewhere, and $\oplus$ is the exclusive OR operator.
Thus, one may regard that $N$ bits of the bit string as additional qubits and $X_a$ as the Pauli $X$ operator on them.

For each pair $i,j$ of computational basis states of $\cH$, choose randomly some permutation $\pi_{i,j}(\cdot)$ of the bit strings of length $N$.
Here we emphasize that this
permutation is chosen from a set of size $(2^N)!$, i.e. permutations of a set of size $2^N$, rather than the much smaller set of size $N!$ of permutations of individual bits in a string.
Choose these permutations uniformly and independently subject to the condition that $\pi_{j,i}$ is the inverse function of $\pi_{i,j}$ and subject to choosing $\pi_{i,i}$ to be the identity permutation for all $i$.
Define $H'$ by
\be
\label{Hprimedef}
H'=-T \sum_a X_a +\sum_{i,j,x} |i,x\rangle \langle j,\pi_{i,j}(x)| \Bigl( \langle i | H | j \rangle \Bigr),
\ee
where $T>0$ is a scalar and the second term, in words, means that for each pair $i,j$ of computational basis states of $\cH$, for each $x$, if there is a matrix element of $H$ between $i$ and $j$, then there is a matrix element of $H'$ between $i,x$ and $j,\pi_{i,j}(x)$.
Note that since $\pi_{i,i}$ is the identity permutation, if $H$ has no sign problem then $H'$ has no sign problem.

Remark: we have written $H'$ using a particular choice of basis states.  However, we can, as in \cref{perminto}, assume that $H'$ is conjugated by a further random permutation so that the algorithm has no information on the labels of the basis states.
In this case, if the algorithm receives some state $|i,x\rangle$ as a neighbor in response to some query, and some other state $|i,y\rangle$ in response to a query with $x\neq y$, the algorithm will be unable to know that in both cases the first index $i$ is the same.

The key idea of this construction, using this remark, is that we can ensure that the algorithm is exponentially unlikely to receive the same label twice in response to a query, except for some trivial situations.

Given a path of Hamiltonians $H_s$, we can define a path $H'_s$ in the obvious way.
Define an isometry $\isL$ from $\cH$ to $\cH'$ by
$$\isL=2^{-N/2} \sum_{i,x} |i,x\rangle \langle i|.$$
Choosing $T>0$, the ground state subspace of $-T\sum_a X_a$ is the range of $\isL$, and if $H_s$ has no sign problem then
for any $T>0$
the ground state of $H'_s$ is equal to
$$\psi'_s=\isL \psi_s,$$
where $\psi_s$ is the ground state of $H_s$. 
To see this, 
let us refer to the second term of \cref{Hprimedef}, i.e. $\sum_{i,j,x} |i,x\rangle \langle j,\pi_{i,j}(x)| \Bigl( \langle i | H | j \rangle \Bigr)$, as the {\it cover} of $H$.  
By Perron-Frobenius, the ground state of the cover of $H_s$ has a nonzero projection onto the range of $\isL$; further the cover of $H_s$ commutes with
the projector $\isL \isL^\dagger$, so the ground state of the cover of $H_s$ is in the range of $\isL$.
So, the ground state of the cover of $H_s$ is $\isL \psi_s$ as claimed.  Hence, for $T>0$, the
ground state of $H'_s$ is as claimed. 
Further,
 if the path $H_s$ is admissible, so is the path $H'_s$, as long as we choose
 $T$ at least inverse polynomially large so that the fifth condition on the spectral gap is satisfied.

The path $H'_s$ does not yet satisfy the endpoint condition.  However, this is easy to resolve if the path $H_s$ obeys the endpoint condition.  
Concatenate the path $H'_s$ with a final path along
which the term $-T \sum_a X_a$ is replaced by the more general term
$$K_{\theta}\equiv -T\Bigl( \cos(\theta) \sum_a X_a + \sin(\theta) \sum_a Z_a\Bigr),$$
with $Z_a$ being the Pauli $Z$ operator on the additional qubits, i.e., $Z_a=\sum_{i,x} (1-2x_a)  |i,x\rangle \langle i,x|,$
where $x_a$ is the $a$-th entry of bit string $x$.
On this final path, vary from $\theta=0$ to $\theta=\pi/2$.
Similarly, also concatenate with an initial path along which we vary from $\theta=\pi/2$ to $\theta=0$.
Let $\tilde H_s$ denote the path of Hamiltonians given by $H'_s$ concatenated with these two additional paths.
Then, if the ground state of $H_0$ is $|i\rangle$, the ground state of the $\tilde H_s$ at the start of the path is given by $|i,0\rangle$, and similarly at the end of the paths.

Further, concatenation with these additional paths preserves the spectral gap.
Let the ground state of $H_0$ be $|i\rangle$.  Define a $2^N$-dimensional subspace spanned by states of the form $|i,x\rangle$ over all $x$.
The Hamiltonian on the additional initial path does not couple this subspace to the orthogonal subspace.  It is easy to compute the spectrum in this subspace,
since the Hamiltonian in this subspace is just a $\langle i | H_0 | i \rangle+K_\theta$.  One may verify the gap in this subspace, and verify that the ground state energy in this subspace is
$\langle i | H_0 |i\rangle$ plus the minimum eigenvalue of $K_\theta$. Any state orthogonal to this subspace must have 
energy at least equal to the minimum eigenvalue of 
$K_\theta$ plus the smallest eigenvalue of
$H_0$ in the space orthogonal to $|i\rangle$ (i.e., at least the second smallest eigenvalue of $H_0$) so the gap follows.
A similar argument holds for the final path.

Thus, we have an admissible path of Hamiltonians $\tilde H_s$ satisfying the endpoint condition, with the ground state of $\tilde H_s$ trivially related to that of $H_s$ for $s=0,1$ (the polynomially small error in $\psi'_s=\isL \psi_s$ at intermediate steps of the path is unimportant for this).

Suppose now we give the algorithm some additional information in response to queries: if we query some state $|i,x\rangle$ and a neighbor is some other state $|i,y\rangle$, then the algorithm will be informed that the value of $i$ remains the same.  This additional information can only help.
However, we claim that with this additional information, {\it up to exponentially small error, the queries of $\tilde H_s$ in the original query model can be described by queries of $H_s$ in the modified query model}.
More precisely, assume we know that $\tilde H_s$ is given by this construction.  Then, the only information given by querying $\tilde H_s$ along the ``final" or ``initial" paths above where $\theta$ varies is that one may get multiple labels which are known to have the same first index, i.e., one may start with $|i,0\rangle$, and then get labels describing other states $|i,x\rangle$.  Further, we claim that up to exponentially small error explained below, queries along the path $H'_s$ in the original query model can be described by queries of $H_s$ in the modified query model.

To show this,
consider the probability that the algorithm receives the same label twice in response to a query.
Suppose the 
algorithm makes multiple queries in which the first label does not change and the algorithm knows it due to the additional information above.  Thus, the algorithm will know that some set of labels will describe the same value of the first index.
After some number of queries, there will be several sets $S_1,S_2,\ldots$ where each set is a set of labels known to describe the same value of the first index.  Formally, there is an equivalence relation on labels: two labels are equivalent if one label is received in response to a query on the other and it is known that the first index did not change, and we extend this equivalence transitively, and the sets $S_1,S_2,\ldots$ are equivalence classes under this relation.
Now, consider the probability that some query of some label $|i,x\rangle$ gives a label $|j,y\rangle$ that has been seen previously by the algorithm in response to a previous query, in the case that $j\neq i$ (so that this query does not simply increase the size of one of the equivalence classes, but actually yields new information about the Hamiltonian).
The second index $y$ equals $\pi_{j,i}(x)$ and $\pi_{j,i}(x)$ is chosen uniformly at random subject to the condition that $\pi_{j,i}$ is the inverse of $\pi_{i,j}$.  
Hence, after only quasi-polynomially many queries (so that only quasi-polynomially values of $\pi_{j,i}$ have been fixed) it is exponentially unlikely that $\pi_{j,i}(x)$ will agree with any previously given value of the second index, unless it is the case that
we have previously queried $|j,\pi_{j,i}(x)\rangle$, i.e., unless $|i,x\rangle$ was received as a label of a neighbor of $|j,\pi_{j,i}(x)\rangle$, which is precisely the case in the modified query model that we receive the same label for a given value of the
state.  Now consider a query in which the first index does not change;
suppose we queried a vertex in some equivalence class $S$, receiving some new label $|i,x\rangle$.   It is exponentially unlikely that this label labels a state in some other equivalence class, though it may be only polynomially unlikely that it labels a state in the given class $S$.  Hence, except for an exponentially small probability, a query in which the first index does not change will not collapse two different equivalence classes.

Hence we have:
\begin{lemma}
If  \cref{mainth} holds in the modified query model, then it holds in the original query model.
\end{lemma}

\section{Distinguishing Graphs}
\label{distg}
Now, within the modified query model we show how to use two (families of) graphs $C,D$ to construct instances of $\Ocl$ to prove  \cref{mainth}.  The main result is \cref{blemma}, which we give at the end of this section after developing the machinery of paths needed.
In this section we assume that several properties of $C,D$ hold.  We summarize these in \cref{trq}.  More detail is given below and these properties are proven in later sections of the paper

Both graphs $C,D$ will have a privileged vertex called the ``start vertex".  For tree graphs, the start vertex will often be the root of the tree.  

Given a graph $G$, we say that the Hamiltonian corresponding to that graph is equal to minus the adjacency matrix of that graph, where each vertex of the graph corresponds to a distinct computational basis state.
We will assume that the ground state energy of the Hamiltonian of $C$ is lower than the ground state energy of $D$ by a spectral gap that is $\Omega(1/\poly(N))$; indeed, the difference will be much larger than that in our construction.
We will also assume that the gap between the ground state of the Hamiltonian of $C$ and the first excited state of that Hamiltonian is also $\Omega(1/\poly(N))$; indeed, that difference is also much larger than that.
Further, we will assume that the amplitude of the ground state wavefunction of $C$ on the start vertex is also $\Omega(1/\poly(N))$.

At the same time, we will also assume a superpolynomial lower bound on the number $q$ of classical queries needed to distinguish $C$ from $D$ in the modified query model above, assuming that the first vertex queried is the start vertex.
The modified query model refers to querying a Hamiltonian; here the Hamiltonian will be the Hamiltonian of the given graph, so that computational basis states are neighbors if the corresponding vertices are neighbors.
The bound is given in terms of mutual information in \cref{distlemma}; if the algorithm is randomized, then the mutual information is conditioned on any randomness used by the algorithm.
Later choices of constants in \cref{chs} will make $q=\exp(\Theta(\log(N)^2))$.

\begin{center}
\begin{itemize}
\item[1.] The ground state energy of the Hamiltonian of $C$ lower than that of $D$ by $\Omega(1/\poly(N))$.

\item[2.] $\Omega(1/\poly(N))$ gap of Hamiltonian of $C$.

\item[3.] Amplitude of ground state of $C$ on start vertex is $\Omega(1/\poly(N))$.

\item[4.] Lower bound on number of classical queries to distinguish $C$ from $D$ in the modified query model.  Precisely: if the graph is chosen randomly to be $C$ with probability $1/2$ and $D$ with probability $1/2$, then with fewer than $q$ classical queries the mutual information (in bits) between the query responses and the choice of graph is bounded for all sufficiently large $N$ by some quantity which is strictly less than $1$, for some $q$ which is superpolynomial.

\item[5.] Number of vertices is $O(2^{\poly(N)})$.  This property is needed because each vertex will correspond to some computational basis state.
\end{itemize}
\captionof{table}{List of properties needed for graphs $C$ and $D$.} \label{trq}
\end{center}

Then, given these graphs, we now construct a path of Hamiltonians $H_s$.  To describe this path, we start with a simplified case.  Consider a problem where for every vertex of a graph $G$ there is a corresponding
computational basis state, and there is one additional computational basis state written $|0\rangle$ which is distinct from all the basis states corresponding to vertices of the graph.  We will label the basis state corresponding to the start vertex of $G$ by $|s\rangle$ (hopefully no confusion will arise with the use of $s$ as a parameter in the path).
Consider the two parameter family of Hamiltonians
\be
\label{HtU}
H(t,U)=-U|0\rangle\langle 0|+t\Bigl( |0\rangle\langle s| +h.c.\Bigr)+H(G),
\ee
where $H(G)$ is the Hamiltonian corresponding to the graph $G$ and where $U,t$ are both scalars\footnote{The terminology $U$ is used as it is a commonly used notation in physics for such diagonal terms; hopefully no confusion arises with the use of $U$ for unitaries in quantum information.}.
We take $t<0$ so that the Hamiltonian has no sign problem.

Now consider a path of Hamiltonians starting at very negative $U$ and with $t=0$ (so that initially the ground state is $|0\rangle$ for both $C$ and $D$),
 then making $t$ slightly negative and increasing $U$ to a value between the ground state energy of 
 $C$ and that of $D$, followed by returning $t$ to zero.  
 At the
end of this path, if $G=D$, the ground state of the Hamiltonian will still be $|0\rangle$ but if $G=C$, the ground state of the Hamiltonian will be the ground state of $H(C)$.
We make $t$ nonzero during the middle of the path to avoid a level crossing: if $t=0$ throughout the path, then the gap closes if $G=C$ when $U$ becomes equal to the ground state energy of $C$.

Indeed, we may choose the path dependence of parameters $U,t$
so that we get an admissible path assuming the properties of $C$ and $D$ above.
For both $C$ and $D$, the ground state energy of $H(G)$ is only $\poly(N)$ so we may take $U$ only polynomially large initially. 
Giving the rest of the path in detail,
change $t$ to an amount $-\Omega(1/\poly(N))$ and then change $U$ so that it is $1/\poly(N)$ larger than the ground state energy of $H(C)$, but still much smaller than the ground state energy of $H(D)$ and than the first excited state energy of $H(C)$.  Do this with $|t|$ much smaller than the spectral gap of $H(C)$ and much smaller than the difference between the ground state energy of
$H(C)$ and the ground state energy of $H(D)$; given the differences in ground state energies of $H(C)$ and $H(D)$ and the gap of $H(D)$, it is possible to do this with $|t|$ that is indeed $\Omega(1/\poly(N))$ so that the gap of the Hamiltonian $H(t,U)$ remains $\Omega(1/\poly(N))$.  Finally return $t$ to $0$.

For use later, let us call the path defined in the above paragraph $P(G)$.

At first sight, this path $P(G)$, combined with the lower bound {\bf 4} of \cref{trq} might seem to solve the problem of the needed separation between problems in $\Ocl$ and the power of classical algorithms: the classical algorithm cannot distinguish the two graphs but one can distinguish them with an admissible path of Hamiltonians.  However, this is not true; for one, our path of Hamiltonians does not satisfy the endpoint condition as the ground state of the Hamiltonian at the end of the path is a superposition of basis states in the case that $G=C$.  Further, the problem is to compute the basis vector at the end of the path, not to distinguish two graphs\footnote{Remark: we could have defined a different oracle problem which differs from $\Ocl$ in two ways: first, we only require that the admissible path satisfy the endpoint condition at endpoint $s=0$ and second, we say that an algorithm ``solves" the problem if it computes the amplitude of the basis state $|i\rangle$ which is the ground state of $H_0$ in the wavefunction which is the ground state of $H_1$ to within error $1/\poly(N)$ with success probability $\geq 2/3$.  Then, for graph $C$ this amplitude is $0$ and for $D$ this amplitude is $1$ but the classical algorithm cannot distinguish the two cases with probability much larger than $1/2$.  If we considered this class, then the given path would prove the needed separation.  However, this is not what we are considering.}.

We might try to solve this by concatenating the path of Hamiltonians above with an additional path that decreases $H(G)$ to zero while adding a term $V |s\rangle\langle s|$ and with $V$ decreasing from zero, so that the ground state of the Hamiltonian (in the case that $G=C$) evolves from being the ground state of $H(G)$ to being $|s\rangle$ and the path now satisfies the endpoint condition.

This still however does not solve the problem: for this path $H_s$, a classical algorithm can determine the ground state at the end of the path (which we will assume to occur at $s=1$) by querying the oracle three times, first querying $\langle 0 | H_1 | 0\rangle$, then querying the oracle to find the neighbors of $|0\rangle$ in the middle of the path (so that it can determine $s$), and
finally querying $\langle s | H_1 | s \rangle$.

So, to construct the path showing  \cref{mainth}, we use an additional trick.
First, we consider $N$ different copies of the problem defined by Hamiltonian \cref{HtU} ``in parallel".  Here, taking ``copies in parallel" means the following, given several Hamiltonians $H_1,H_2,\ldots,H_N$, with associated Hilbert spaces $\cH_1,\cH_2,\ldots,$, we define a Hamiltonian $H$ on Hilbert space $\cH\equiv \cH_1 \otimes \cH_2 \otimes \ldots$ by
$H=H_1\otimes I \otimes I \ldots+ I\otimes H_2 \otimes I \otimes \ldots + \ldots$ where $I$ is the identity matrix.
There is an obvious choice of computational basis states for $\cH$, given by tensor products of computational basis states for $\cH_1,\cH_2,\ldots$.
Similarly, given paths of Hamiltonians, we consider the paths in parallel in the obvious way.
If each path is admissible, then the path given by those paths in parallel is also admissible; note that the size of the Hilbert $\cH$ is $2^{\poly(N)}$ if each $\cH_i$ has dimension $2^{\poly(N)}$.

For each of these $N$ copies, we choose $G$ to be either $C$ or $D$ independently, so that there are $2^N$ possible instances.
Write $G_i$ to denote the graph chosen on the $i$-th copy.
For each copy $i$ let $P_i$ denote the path $P$ for that copy.  Let $\tilde P$ denote the path given by taking all those paths $P_i$ in parallel.  At the end of the path $\tilde P$, the ground state is a tensor product of $|0\rangle$ on some copies and the ground state of $H(C)$ on some other copies.

To give an intuitive explanation of the trick we use, it will be that we use this property of the ground state as a kind of key to find an entry in a database: there will be some projector ($\Pi$ below) which is diagonal in the computational basis and one must find entries of it which are nonzero, and the adiabatic evolution will use this property of the ground state to find it, but the classical will not be able to.

The trick is: concatenate that path $\tilde P$ with a further path $Q$.  To define this path $Q$, add two additional terms to the Hamiltonian
$$V\sum_i |s\rangle_i\langle s| +W \Pi,$$
where $|s\rangle_i\langle s|$ denotes the projector onto $|s\rangle$ on the $i$-th copy tensored with the identity on the other copies, where $V,W$ are scalars, and where $\Pi$ is a projector which is diagonal in the computational basis.
The projector $\Pi$ will be equal to $1$ on a given computational basis state if and only if that computational basis state is a tensor product of basis state $|0\rangle_i$ on all copies for which $G_i=D$ and of basis states corresponding to vertices of $G_i$ for all copies on which $G_i=C$.
Thus, the ground state at the end of path $\tilde P$ is an eigenvector of $\Pi$ with eigenvalue $1$.
The  path $Q$ is then to first decrease $W$ from zero so that it is large and negative (it suffices to take it polynomially large) while keeping $t=0$; then decrease the coefficient in front of $H(G)$ to zero while increasing $V$ to be $\Omega(1/\poly(N))$.  

Making $W$ large and negative ensures that throughout $Q$,
the ground state of the Hamiltonian is in the eigenspace of $\Pi$ with unit eigenvalue.
This decrease in the coefficient in front of $H(G)$ combined with increase in $V$ ensures that the ground state at the end of the path is a computational basis state: it is a tensor product of 
$|0\rangle_i$ on all copies for which $G_i=D$ and of states $|s\rangle_i$ for all copies on which $G_i=C$.
We choose $V$ to be $\Omega(1/\poly(N))$ so that the gap of the Hamiltonian will be $\Omega(1/\poly(N))$.

Let $\hat P$ be the concatenation of $\tilde P$ and $Q$.  Note that there are $2^N$ possible instances of path $\hat P$, depending on different choices of $G_i$.

Now we bound the probability of a classical algorithm to determine the final basis state. 
This lemma shows that if we construct graphs which satisfy {\bf 1-5} of \cref{trq}, then \cref{mainth} follows. 
\begin{lemma}
\label{blemma}
If items {\bf 1-3,5} of \cref{trq} hold, then $\hat P$ is an admissible path.
Further, if item {\bf 4} holds,
 no algorithm 
 which uses a number of classical queries which is quasi-polynomial and is smaller than $q$
can solve all instances with probability greater than $\exp(-cN)$.
\begin{proof}
By construction $\hat P$ is admissible.

Choose each $G_i$ independently, choosing it to be $C$ with probability $1/2$ and $D$ with probability $1/2$.

By item {\bf 4} of \cref{trq}, with fewer than $q$ queries of the oracle the mutual information\footnote{Here, the random variables considered are as follows.  First, a single bit for each graph $G_i$ to determine whether the graph is $C$ or $D$.  Then, additional randomness to determine the labels in the modified query model.  Then, any randomness used by the algorithm to choose queries (if the algorithm is randomized we then consider the mutual information conditioned on that randomness, and similarly we consider entropies conditioned on that randomness later in the proof).  Finally,
the query responses (which of course are determined by the other random variables).} in bits between each $G_i$ and the query responses is bounded by some quantity $S<1$.
Of course, if the algorithm makes less than $q$ queries, then the average number of queries per graph is even less (i.e., less than $q/N$), but certainly for any $i$, we make at most $q$ queries of graph $G_i$ and so the mutual information between any $G_i$ and the set of all query responses is at most $S<1$.

Hence, the average entropy of $G_i$ given the query responses is at least $1-S$,
as $G_i$ is a random variable with entropy $1$.  
 So, since the entropy of $G_i$ given the query responses is bounded by $1$, with probability at least $(1-S)/2/(1-(1-S)/2)$
the entropy of $G_i$ given the query responses is at least $(1-S)/2$.

Thus, with probability that is $\Omega(1)$, the entropy of $G_i$ is $\Omega(1)$.
To get oriented, assume that these events (the entropy of each $G_i$) are independent; that is, define $N$ additional binary random variables $S_i$ which quantify the entropy of $G_i$ being $\geq (1-S)/2$ or not, and assume that these are all independent.
Then, with probability $1-\exp(-\Omega(N))$, there are $\Theta(N)$ independent variables $G_i$ each of which\footnote{Indeed, Bayes' theorem implies that the $G_i$ are independent variables given the query responses.  The prior distribution is to choose the $G_i$ uniformly and independently, so they are independent variables in the prior.  The likelihood of getting a certain set of responses in response to a given set of queries is a product over graphs of a likelihood function for each graph (with that function determined just by the queries of that graph).  So, they are independent variables in the posterior.}
 have entropy $\Omega(1)$ and
so it is not possible to determine all $G_i$ with probability better than $\exp(-cN)$ using only quasi-polynomially many queries.

Finally, consider the possibility that the $S_i$ are not independent.  For example, there is a rather silly algorithm that makes these $S_i$ dependent on each other: consider any algorithm $A$ that gives independent $S_i$ and define a new algorithm $A'$ that calls $A$ with probability $1/2$ and makes no queries with probability $1/2$ (in which case all $S_i=1$ since no information is known about any $G_i$).  Then the variables $S_i$ for $A'$ are not independent.  However, this ``silly algorithm" certainly does not help.

Still we must consider the possibility that there is some way of correlating the $S_i$ which would help.  
Suppose there were an algorithm which gave correlated $S_i$, so that for some $i$
the mutual information between $G_i$ and query responses,
conditioned on some responses for $j\neq i$ and conditioned on the $G_j$ for $j\neq i$, was larger than $S$.  However, we could then postselect this algorithm on the query responses to the set of $j \neq i$ to give an algorithm that just acted on copy $i$ but which gave mutual information greater than $S$.
\end{proof}
\end{lemma}

\section{A First Attempt}
\label{toomany}
In this section we give a first attempt at constructing two graphs, $C$ and $D$ which satisfy the properties of \cref{trq}.
Unfortunately, the example will not quite satisfy the fourth property (using the privileged start vertex it will be possible to efficiently distinguish them, but it will not be possible without that knowledge) or
for the fifth: the graph $D$ will have too many vertices.
The construction later will fix both of these defects.

Briefly, the graph $C$ is a complete graph on $4$ vertices, i.e., every one of the four vertices has degree $3$ so it connects to every other vertex.
The graph $D$ is a tree graph where every vertex except the leaves has degree $3$, i.e., $D$ is given by attaching three binary trees to some root vertex.  
We choose all the leaves of $D$ to be at distance $h$ from the root for some $h$ so that $D$ has $1+3+2\cdot 3 + 2^2\cdot 3+\ldots+2^{h-1} \cdot 3$ vertices.

We choose any vertex of $C$ arbitrarily to be the start vertex.  We choose the start vertex of $D$ to be the root.  Then, it is trivial to verify item {\bf 1-3} of \cref{trq}.

Consider what it means to distinguish two graphs in the modified query model.  A query of any vertex can return only the information of the degree of that vertex and whether or not that vertex is the start vertex.  Since all matrix elements between the computational basis state of that vertex and its neighbors are the same, one cannot distinguish the different neighbors in any way from the response to the given query.  As we have mentioned, though, one can determine if a queried vertex is the start vertex since the Hamiltonian will have an additional coupling
$t\Bigl( |0\rangle\langle s| +h.c.\Bigr)$.

Using the knowledge of which vertex is the start vertex it is not hard to distinguish the two graphs in $O(1)$ queries: query the start vertex $s$ to get some new vertex $v$, then query a neighbor of $v$ (other than $s$, i.e. nonbacktracking) to get some new vertex $w$, then finally query a neighbor of $w$, again without backtracking.  For $C$, with probability $1/3$ that neighbor of $w$ will be the start vertex.  On the other hand, for $D$, the neighbor will never be the start vertex.

Suppose however, that we use only the information about the degree of the vertex and not which vertex is the start vertex.
In this case, it is impossible to distinguish $C$ from $D$ using fewer than $h$ queries because, trivially, any vertex accessed with fewer than $h$ queries is not a leaf of $D$ and hence has the same degree (i.e., $3$) as every vertex in $C$, and hence the terms in the Hamiltonian coupling the corresponding computational basis state to its neighbors are the same.

So, if we choose $h$ superpolynomially large, then item {\bf 4} is ``almost satisfied", i.e., without information about the start vertex we cannot distinguish them with polynomially many queries.
However, if we choose $h$ superpolynomially large, then item {\bf 5} is not satisfied since the number of vertices is not $O(2^{\poly(N)})$.

If instead we choose $h$ only polynomially large,
there is a simple efficient classical algorithm to distinguish $C$ from $D$ even without using knowledge of the start vertex: since we can avoid backtracking in the given query model, we will arrive at a leaf in $h$ queries.  Even if we perform a random walk on $D$, allowing backtracking, we will typically arrive at a leaf in $O(h)$ queries

\section{Decorated Graphs}
\label{decg}
In this section we define an operation called {\it decoration} and then define graphs $C,D$ in terms of this operation.  Our decoration operation is very similar to (indeed, it is a special case of) the decoration
operation defined in \cite{Schenker_2000} (there are other uses of the term ``decoration" in the math literature, such as in set theory, which are unrelated to this).
 These graphs $C,D$ will depend upon a large number of parameters; in \cref{chs} we will show that for appropriate choice of these parameters, all properties in \cref{trq} are fulfilled.
 
 \cref{decdef} defines decoration.  
 The idea of decoration is to make it easy for a classical algorithm to ``get lost".  Decoration will add additional vertices and edges to some graph, and a classical algorithm will tend to follow what one may call ``false leads" along these edges so that it is hard for it to determine properties of the graph before decoration because it takes a large number of queries to avoid these false leads.
  
  \cref{applic} applies decoration to define $C,D$ and explains some motivation for this choice of $C,D$.  \cref{spec} considers the spectrum of the adjacency matrices of $C,D$, as well as proving some properties of the eigenvector of the adjacency matrix of $C$ with largest eigenvalue.  Here we, roughly speaking, bound the effect of decoration on the spectrum and leading eigenvector of the graph.
 
\subsection{Decoration}
\label{decdef}
In this subsection, we define an operation that we call {\it decoration} that maps one graph to another graph.

We first recall some graph theory definitions.  For us, all graphs are undirected, so all edges are unordered pairs of vertices, and there are no multi-edges or self-edges so that the edge set is a set of unordered pairs of distinct vertices.
An $m$-ary tree is a rooted tree (i.e., one vertex is referred to as the root) in which each vertex has at most $m$ children.  A full $m$-ary tree is a tree in which every vertex has either or $0$ or $m$ children (i.e., all vertices which are not leafs have $m$ children).
A perfect $m$-ary tree is a full $m$-ary tree with all leaves at the same distance from the root.  A binary tree is an $m$-ary tree with $m=2$.

We will  make an additional definition (this concept may already be defined in the literature but we do not know a term for it).  We will say that a
graph is ``$d$-inner regular with terminal vertex set $T$" if $T$ is a set of vertices
such that every vertex not in $T$ has
degree $d$.  The vertices in $T$ will be called terminal vertices and the vertices not in $T$ will be called inner vertices.

We define the height of a tree to be the length of the longest path from the root to a leaf, so that tree with just one vertex (the root) has height $0$ (sometimes it is defined this way in the literature, but other authors define it differing by one from our definition).
Then, the number of vertices in a perfect $m$-ary tree of height $h$ is
$1+m+m^2+\ldots+m^{h}=(m^{h+1}-1)/(m-1)$.

We now define decoration.  If we $(c,m,h)$-decorate a graph $G$, the resulting graph is given by attaching
$c$ perfect $m$-ary trees of height $h-1$ to each vertex of $G$; that is, for each vertex $v$ of $G$, we add $c$ such trees, adding one edge from $v$ to the root of each tree, so that the degree of $v$ is increased by $c$.
Call the resulting graph $H$.
Then, any vertex of graph $G$ which has degree $d$ corresponds to a vertex of $H$ with degree $d+c$.
We may regard $G$ as an induced subgraph of $H$; if $G$ is a rooted tree, then the root of $G$ corresponds to some vertex in $H$ that we will regard as the root of $H$.
We will refer to those vertices of $H$ which correspond to some vertex of $G$ as the {\it original vertices} of $H$, i.e., the original vertices of $H$ are those in the subgraph $G$.
If $G$ is a full $n$-ary tree, and if we $(n,2n,h)$ decorate $G$, then $H$ is a full $(2n)$-ary tree.

Now we define a sequence of decorations.  Consider some sequence of heights $h_1,h_2,\ldots,h_l$ for some $l$.  Given a graph $G_0$ which is $(n+1)$-inner regular, we $(n,2n,h_1)$-decorate $G$, calling the result $G_1$.  
The terminal set of $G_1$ will be the set of vertices which correspond to vertices in the terminal set of $G$, as well as additional leaves (vertices of degree $1$) added in decoration, so that $G_1$ is $(2n+1)$-inner regular.
We then $(2n,4n,h_2)$-decorate $G_1$, calling the result $G_2$, defining the terminal set of $G_2$ in the analogous way.
Proceeding in this fashion, we $(2^m\cdot n,2^{m+1}\cdot n,h_m)$-decorate $G_m$, giving $G_{m+1}$, until we have defined graph $G_l$.  We say that $G_l$ is given by decorating $G$ with height sequence $h_1,\ldots,h_l$.
Note that each graph $G_m$ is inner regular.

We will call the {\it original vertices} of $G_l$ those vertices which correspond to some vertex of $G_0$ in the obvious way, i.e., $G_0$ is an induced subgraph of $G_k$ and the original vertices are the vertices in that subgraph.

Remark: the reason that we talk about $(n+1)$-inner regular, rather than $n$-inner regular is that if one avoids backtracking, this means that there are $n$, rather than $(n-1)$, choices of vertex to query from any given vertex.  
Also, since we are decorating with trees, in this way $n$ refers to the arity of the tree rather than the degree of the tree.
This is just an unimportant choice of how we define things and other readers might find it more convenient to shift our value of $n$ by one.
 
For use later, let us define 
$T_{n,h_0}$ where $T_{n,h_0}$ is constructed by attaching $n+1$ perfect $n$-ary trees to some given vertex, i.e., $T_{n,h_0}$ is $n$-inner regular with the terminal set being the leaves of the trees.

\begin{figure}
\includegraphics[width=3in]{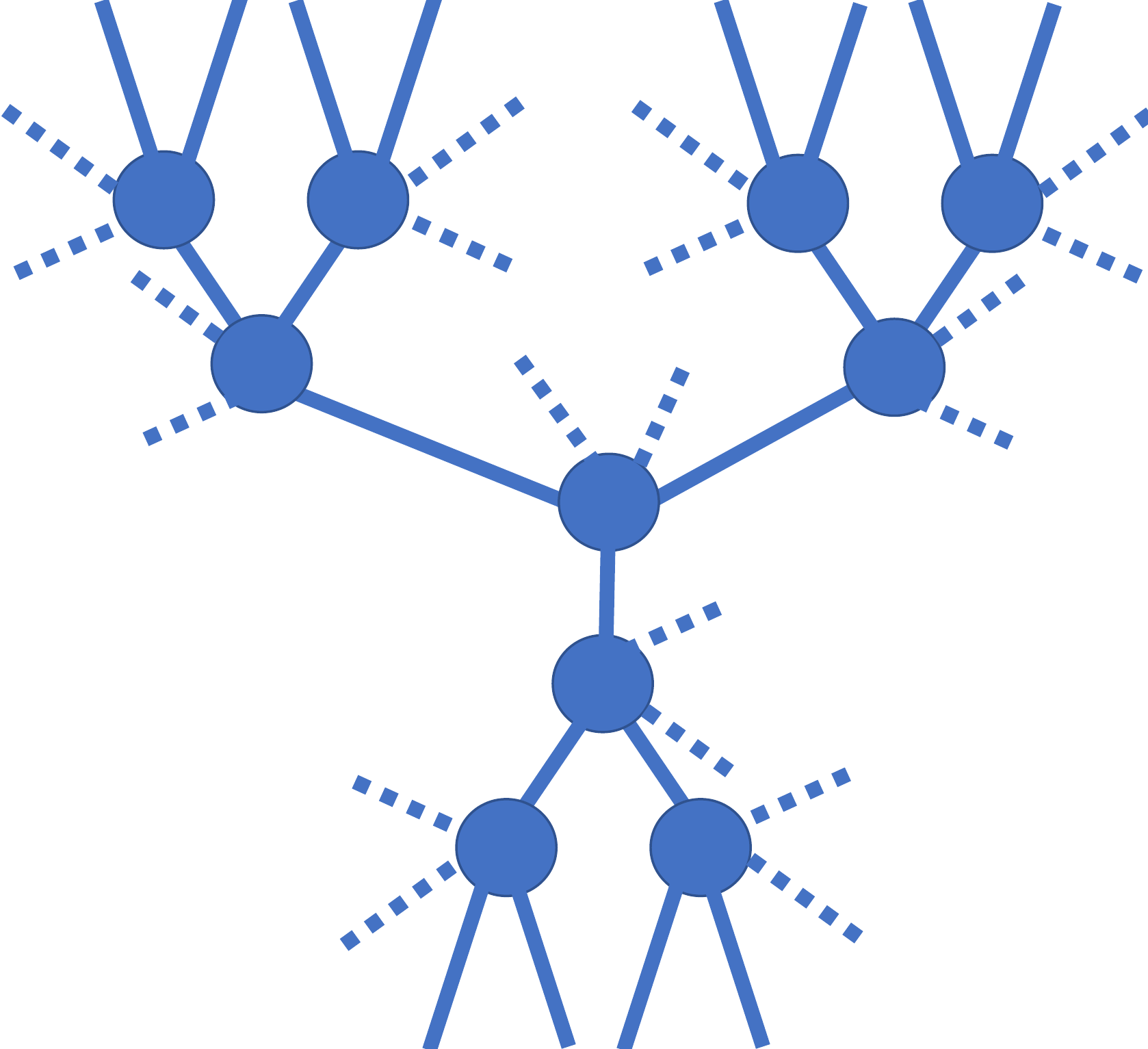}
\caption{Example of decoration.  Solid lines are edges of graph $G$ which is $3$-inner regular.
In the notation given, it is $T_{3,2}$.  Dashed lines represent edges added after $(2,4,1)$-decorating $G$.  Solid circles are inner vertices and terminal vertices are not shown.
Since $h_1=1$, the $4$-ary trees added are of height $h_1-1=0$, consisting just of the root; if we had taken $h_1=2$, then each of the dashed lines shown would have a solid circle and four additional dashed lines attached to them.}
\label{figdec}
\end{figure}

See Fig.~\ref{figdec} for an example of decoration of such a graph $T_{3,2}$.

\subsection{Application of Decoration, and Universal Cover of Graph}
\label{applic}
We will apply this decoration to two different choices of $(n+1)$-inner regular graphs, each of which has some fixed vertex that we call the start vertex.

In the first case, we pick $G_0$ to have vertices labelled by a pair of integers $(i,j)$ with
$0\leq i <\glen$ for some $\glen>1$ and $0\leq j <m$ for $m= (n+1)/2$ for some odd $n$.
There is an edge between $(i,j)$ and $(k,l)$ if $i=k\pm 1 \mod \glen$; note that there is no constraint on $j,l$.
So, for $m=1$, the graph is a so-called ring graph.
The start vertex will have $i=\lfloor \glen/2\rfloor $ and $j=0$.
We denote this graph $R_{\glen,m}$ where $m=(n+1)/2$.

The second case is the same except that there is an edge between $(i,j)$ and $(k,l)$ if $i=k\pm 1$.  Note that the ``$\mod \glen$" is missing in the definition of $D$.  
Again the start vertex has $i=\lfloor \glen/2\rfloor $ and $j=0$.
So, for $m=1$, this graph is a so-called path graph or linear graph.
We denote this graph $P_{\glen,m}$.

\begin{figure}
\includegraphics[width=3in]{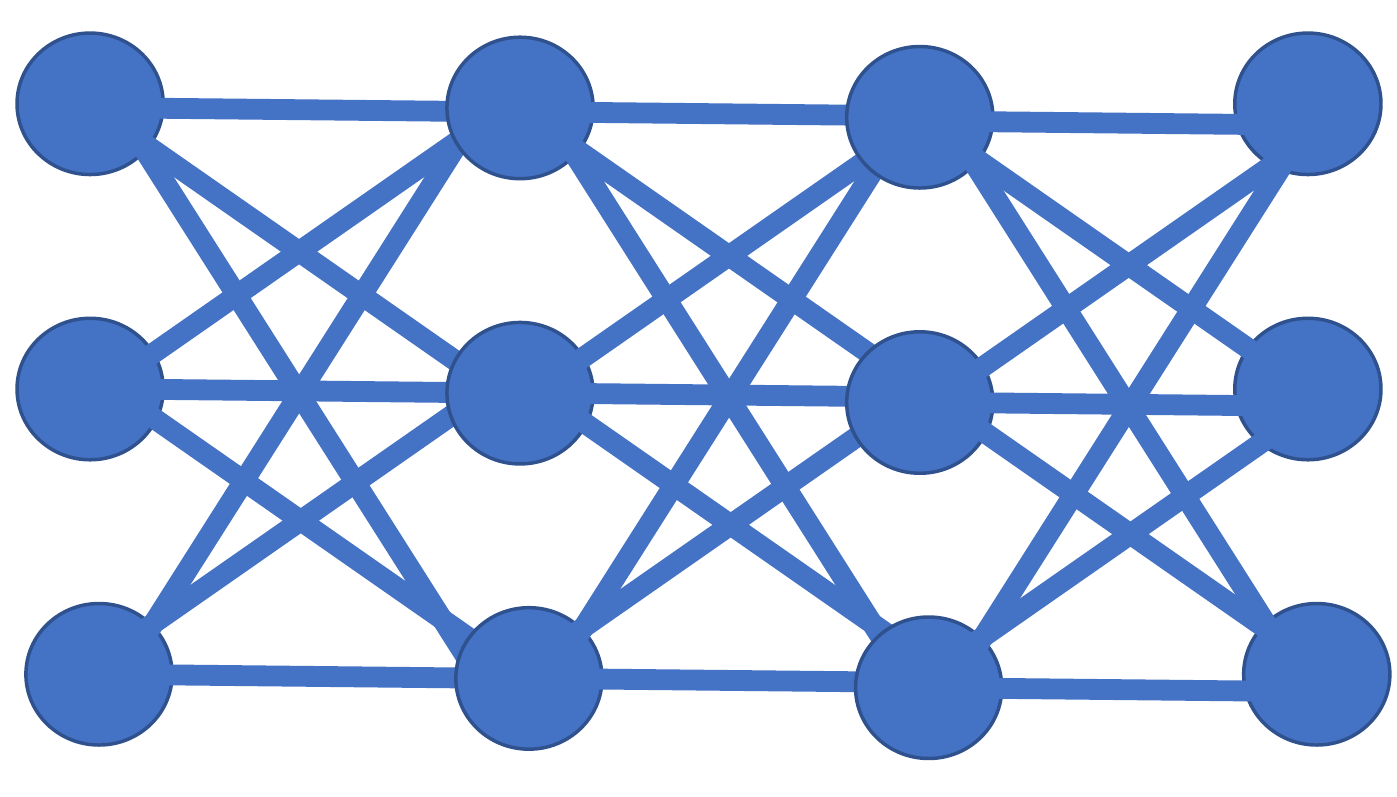}
\caption{Example of graph $P_{4,3}$.  Solid circles are vertices and solid lines are edges.  Note that the graph is $6$-inner regular so $n=5$.  In the case of graph $R_{4,3}$, there are $9$ additional edges connecting each of the
three vertices at the left to each of the three vertices on the right.}
\label{figP}
\end{figure}

See Fig.~\ref{figP}.

Define $C=G_l$ in the case that $G=R_{\glen,(n+1)/2}$ and define
$D=G_l$ in the case that $G=P_{\glen,(n+1)/2}$.

Note that $R_{\glen,(n+1)/2}$ is an $(n+1)$-regular graph and $P_{\glen,(n+1)/2}$ is an $(n+1)$-inner regular graph where the terminal vertices are those $(i,j)$ with $i=0$ or $i=\glen-1$.

Then, in both cases,
the graph $G_l$ is a $d$-inner regular graph with $d=2^{l+1} \cdot n$.
We will define a start vertex on $G_l$ in the obvious way: it is the original vertex in $G_l$ that corresponds to the start vertex of $G$.

The key then is that to distinguish $C$ and $D$, one must be able to go a long distance, of order $L$, on the graph.  Decoration will make it hard for any classical algorithm to follow such a long path.

\subsection{Spectrum of Adjacency Matrix of Decorated Graph}
\label{spec}
We now provide bounds on the spectrum of the adjacency matrix of graph $G_l$ constructed from decorating $G$ with some height sequence.  For any graph $H$, let $\lambda_{0}(H)$ denote the largest eigenvalue of the adjacency matrix of $H$.  Since $G$ is a subgraph of $G_l$, it follows that $\lambda_0(G_l)\geq \lambda_0(G)$.

We have:
\begin{lemma}
\label{eigbounddec}
Assume $n/\glen^2=\omega(1)$.

For either choice of $G_0$, the largest eigenvalue of the adjacency matrix of $G_l$ is bounded by
$$\lambda_0(G)\leq \lambda_0(G_l)\leq \lambda_0(G)+2^{l/2+1} \cdot \sqrt{n}.$$

Further $\lambda_0(C)\geq n$ and
 $\lambda_0(D)\leq n+2^{l/2+1} \cdot \sqrt{n}-\Theta(n/\glen^2)$.

The second largest eigenvalue of the adjacency matrix of $C$ is upper bounded by
$n+2^{l/2+1} \cdot \sqrt{n}-\Theta(n/\glen^2)$.

Finally, let $\psi$ be an eigenvector of the adjacency matrix of $C$ with largest eigenvalue (the eigenspace has dimension $1$ since $G_l$ is connected), with $|\psi|=1$.  
Let $\Pi$ be a diagonal matrix which is $1$ on the original vertices of $C$ and $0$ on the other vertices.
Let $p=|\Pi \psi|^2$.  Remark: heuristically, $p$ is the ``probability" that if one measures $\psi$ in a basis of the vertices, that the result will be one of the original vertices.
Then, if $n \geq 2^{l/2+2} \cdot \sqrt{n}$,
$p\geq 1/5$.
\begin{proof}
To show the first result,
let $A$ be the adjacency matrix of $G_l$.  We decompose $A=A_0+A_1$ where $A_0=\Pi A \Pi$ so that $A_0$ is the adjacency matrix of the subgraph of $G_l$ obtained by deleting all edges except those which connect two original vertices.  Then, $\Vert A_0 \Vert=\lambda_0(G)$ and so $\lambda_0(G_l)\leq \lambda_0(G)+\Vert A_1 \Vert$.

However, $A_1$ is equal to the adjacency matrix of the subgraph of $G_l$ obtained by deleting any edge connecting two original vertices of $G_l$.  This subgraph is a forest, to use the terminology of graph theory: it consists of disconnected trees, one tree for each vertex in $G$.  
Indeed each of these trees is an $m$-ary tree with $m=2^l\cdot n$ (recall that if $G$ has degree $n+1$, then $G_l$ has degree $2^{l}\cdot n+1$).

We now upper bound the largest eigenvalue of the adjacency matrix of an $m$-ary tree.
For a perfect $m$-ary tree of height $h$, it is possible to compute the largest eigenvalue: the tree has a symmetry under permuting the daughters of any given vertex and the largest eigenvector will be invariant under this symmetry.  So, let $v_k$, for integer $k$ with $0\leq k \leq h$, denote the vector with norm $1$ which is an equal amplitude superposition of all vertices which are distance $k$ from the root, so that $v_0$ is $1$ on the root and $0$ elsewhere, $v_1$ has amplitude $1/\sqrt{m}$ on each of the daughters of the root, and so on.  We can then write the adjacency matrix, restricted to the subspace spanned by $v_0,v_1,\ldots$, as
$\sqrt{m} \sum_{0\leq k \leq k+1} |v_k\rangle\langle v_{k+1}|+h.c.$ and so clearly the largest eigenvalue is bounded by $2\sqrt{m}$.
So, $\lambda_0(G_l)\leq \lambda_0(G)+2^{l/2+1} \cdot \sqrt{n},$ as claimed

The lower bound on $\lambda_0(C)\geq n$ follows from the fact that $\lambda_0(R_{\glen,(n+1)/2})=n+1>n$.
The upper bound on $\lambda_0(D)$ follows since $\lambda_0(P_{\glen,(n+1)/2})=n+1-\Theta(n/\glen^2)=n-\Theta(n/\glen^2)$ since $n/\glen^2=\omega(1)$.

To bound the second largest eigenvalue of the adjacency matrix of $C$, again decompose $A=A_0+A_1$.
By the Courant-Fischer-Weyl min-max principle, the second largest eigenvalue of $A$  is equal to the minimum, over all subspaces of codimension $1$, of the largest eigenvalue of the projection of $A$ into that subspace.  Consider the subspace orthogonal to the eigenvector of $A_0$ of largest eigenvalue; the projection of $A_0$ into this subspace is bounded by $n-\Theta(n/\glen^2)$ and so
the second largest eigenvalue of $A$ is upper bounded by $n-\Theta(n/\glen^2)+\Vert A_1 \Vert$.

For the final claim, we have $\lambda_0(C) \leq p \Vert A_0 \Vert + 2 \sqrt{p(1-p)} \Vert A_1 \Vert + (1-p) \Vert A_1 \Vert$.
Since $\lambda_0(C)\geq n$ and $\Vert A_0 \Vert=n$, we have
$(1-p) n \leq \Bigl((1-p)+2\sqrt{p(1-p)} \Bigr) 2^{l/2+1} \cdot \sqrt{n+1}$.
So, for $n\geq 2 (2^{l/2+1} \cdot \sqrt{n})$, after a little algebra we find that $p\geq 1/5$.
\end{proof}
\end{lemma}

\section{Classical Hardness}
\label{ch}
We now show classical hardness.  We will give a lower bound on the number of queries needed by a classical algorithm to distinguish $C$ from $D$, with the initial state of the classical algorithm being the start vertex.  The
lower bound \cref{distlemma} will depend on the difficulty of reaching a certain set $\Delta$ defined below.
To show this difficulty,
the main result is in the inductive \cref{inductivelemma}; we then apply this lemma in \cref{chs}.

Given a $d$-inner regular graph, with some given choice of start vertex, and given some set $S$ which is a subset of the set of vertices, we say that it is $(p,q)$-hard to reach $S$ if no classical algorithm, starting from the start vertex, can reach $S$ with at most  $q$ queries with probability greater than $p$.  Here ``reach" means that at some point the classical algorithm queries a vertex in $S$.

Note that given a perfect $m$-ary tree of height $h$, it is clearly $(0,h+1)$-hard to reach the set of leaves of the tree if the root vertex is the start vertex: the first query will query the start vertex, the second query can query a vertex at most distance $1$ from the start vertex, and so on.

Our goal will be to show for graph $D$ that it is hard (for some choice of parameters $p,q$) to reach the set $\Delta$ of vertices of $D$ which correspond to terminal vertices of $G_0$, i.e., $\Delta$ is a subset of the terminal vertices of $G_l$, consisting only of those which correspond to terminal vertices of $G_0$ rather than leaves added in decoration.

This will then imply a bound on the ability of a classical algorithm to distinguish $C$ from $D$ in the modified query model:
\begin{lemma}
\label{distlemma}
Suppose it is $(p,q)$-hard to reach the set $\Delta$ above for graph $D$.
Suppose one chooses
some graph $G$
to be $C$ or $D$ with probability $1/2$ for each choice.  Then, no classical algorithm using at most $q$ queries in the modified query model can correctly guess which graph $G$ is with probability greater than $1/2+p/2$.

Additionally, the mutual information between the random variable $G$ and the query responses is $\leq p$.
If the algorithm is randomized, the mutual information here is conditioned on the randomness used by the algorithm.
\begin{proof}
Consider the set of vertices in $C$ which correspond to an original vertex $(i,j)$ with $i=0$ or $i=\glen-1$.  In an overload of notation, let us also call this set $\Delta$.

Consider some classical algorithm. 
 If we apply this algorithm to graph $D$ or graph $C$, in the modified query model the two algorithms have the same probability distribution of query responses conditioned on the case that the algorithm applied to $D$ does not reach $\Delta$.

Hence, it is $(p,q)$-hard to reach the set $\Delta$ in $C$.

So, for graph $D$, the probability distribution of query responses is $(1-p) \sigma+p \tau$ for some probability distribution $\tau$: the first term in the sum is the case conditioned on not reaching $\Delta$ and the second is the case when it reaches $\Delta$.
For $C$, the probability distribution of query responses is $(1-p)\sigma+p\mu$ for some probability distribution $\mu$.

The $\ell_1$ distance between the two probability distributions is at most $2p$.  So, the probability that the algorithm guesses right is at most $1/2+p/2$ (to see this, let $P,P'$ be two probability distributions on some set of events; let $e$ label events; choose $P$ or $P'$ with probability $1/2$, then observe some event given the probability distribution; let $A_e=1$ if one choose $P$ on event $e$ and $A_e=0$ otherwise; then the probability of guessing right is $\sum_e A_e P_e/2 - \sum_e (1-A_e) P'_e/2=1/2+\sum_e (A_e-1/2) (P_e-P'_e)/2$ and $\sum_e (A_e-1/2) (P_e-P'_e)/2\leq |P-P'|_1/4$).

The claim about the mutual information follows because the optimal case for the mutual information is when the support of $\sigma,\tau,\mu$ are all disjoint. 
\end{proof}
\end{lemma}

We now need a couple more definitions.
First, 
\begin{definition}
We will assign a 
number called the {\it level} to each vertex $v$ of any graph $G_k$ defined by some sequence of decorations of a graph as follows.
Decoration gives a sequence of graphs $G_0,G_1,\ldots,G_k$.  For $j<k$, we may regard $G_j$ as an induced subgraph of $G_k$.
The level of a vertex $v\in G_k$ is equal to the smallest $j$ such that $v\in G_j$.
Hence, the original vertices of $G_k$ are those with level $0$.
\end{definition}

Second, consider the problem of tossing a coin which is heads with probability $1/2$ and tails with probability $1/2$.
Define $\Pbias(n,N)$ to be the probability that one observes $n$ heads after at most $N$ tosses of the coin.
This is the same as the probability that, after tossing the coin $N$ times, one has observed at least $n$ heads.
Thus,
$$\Pbias(n,N)=\sum_{m=n}^N 2^{-N} {N \choose m}.$$

Now we give a lemma with two parts.  This first part is an inductive lemma that implies difficulty of reaching leaves of a graph obtained by decorating a tree $T_{n,h}$.  The second part is used to prove difficulty of reaching $\Delta$; both parts have very similar proofs and in applications we will first apply the first part inductively for several height sequences and then use the result as an input to the second part of the lemma.
Throughout, we will use $S_{leaf}$ to refer to a set of leafs in a tree graph; this will be the set of terminal vertices.

Now we show:
\begin{lemma}
\label{inductivelemma}
Assume that for some $n,h_0,h_1,\ldots,h_k$, it is $(P,Q)$-hard to reach $S_{leaf}$ in graph $G$ 
given by decorating 
$T_{2n,h_0}$ by height sequence $h_1,\ldots,h_k$.
Then:

\begin{itemize}
\item[{\bf 1.}]
For any integers $M,H\geq 0$, it is $(P',Q')$-hard
to reach $S_{leaf}$ in graph $G'$ given by decorating 
$G=T_{n,H}$ with height sequence $h_0,h_1,\ldots,h_k$,
where
\be
\label{precursion}
P'=\Pbias(H-h_0,M)+Q'P,
\ee
and
\be
Q'=(M-(H-h_0)) Q.
\ee

\item[{\bf 2.}] For any integers $M,\glen\geq 0$,  it is $(P',Q')$-hard to reach $\Delta$ in graph $D$ given by decorating $P_{\glen,(n+1)/2}$ with height sequence
$h_0,h_1,\ldots,h_k$ where
\be
\label{precursion2}
P'=\Pbias(\glen/4-h_0-O(1),M)+Q'P,
\ee
and
\be
Q'=(M-(H-h_0)) Q.
\ee
\end{itemize}

\begin{proof}
We prove the first claim first.
After $q$ queries by the algorithm, we can describe the queries by a tree $T(q)$ with $q$ edges, each vertex of which corresponds to some vertex in $G'$.  We use letters $a,b,\ldots$ to denote vertices in this tree $T(q)$ and use $v,w,\ldots$ to denote vertices in $G'$.
 The root of the tree $T(q)$ corresponds to the start vertex.  If some vertex $a\in T(q)$ corresponds to some $v\in G'$, then the daughters of $a$ correspond to neighbors of $v$ obtained by querying $v$.

Given an $a\in T(q)$, we will say it has some given level if the corresponding vertex in $G'$ has that level.
For any vertex $a\in T(q)$ or any $v\in G'$, we say that the {\it subtree} of $a$  (respectively, $v$) is the tree consisting of $a$ (respectively, $v$) and all its descendants.  ``Queries of a subtree" mean queries in the modified query model starting from the root of that subtree.

Define a subtree $S(q)$ of $T(q)$.  $S(q)$ will be the induced subgraph whose vertices consist of the root of $T(q)$ and of all other vertices $a$ of $T(q)$ which have been queried at least $Q$ times and such that the parent of the given vertex $a$ is at level $0$.

Suppose after some number of queries $q$, some new vertex $a$ is added to $S(q)$.  We will now consider the probability distribution of the level of $a$, conditioned on the level being $0$ or $1$.
Let $a$ correspond to vertex $v$ of $G'$.
Suppose that 
$v$ has distance at most $H-h_0$ from the root of $G'$ so that $S_{leaf}$ in $G'$ is distance at least $h_0$ from $v$.

We will say that a vertex in $G'$ has property $(\dagger)$ if these three conditions hold:
it is distance $\leq H-h_0$ from the root of $G'$, and is level $0$ or $1$, and is the 
the child of a vertex with level $0$.
For a $v$ with property $(\dagger)$, we will say that
event (*) occurs for that $v$ if we reach a vertex in the subtree of $v$ which is level $0$ or $1$ and distance $\geq h_0$ from $v$ with fewer than $Q$ queries in that subtree.
If $v$ has level $1$, then 
the responses to queries of the subtree of $v$ are the same as in the 
given query model on graph $G$ given by
decorating $T_{2n',h_0}$ by height sequence $h_1,\ldots,h_k$.
In that case, event (*) occurs iff we reach $S_{leaf}$ in $G$ in fewer than $Q$ queries; this probability
is bounded by $P$ by the inductive assumption.
If instead
$v$ has level $0$, then since by assumption $S_{leaf}$ in $G'$ is distance at least $h_0$ from $v$, the probability of 
event (*) occurring for that $v$ is also bounded by $P$.
The key point here is that if $v$ has level $0$ and we consider the subtree of $v$, and then further consider the subgraph of that subtree consisting of vertices of distance $\leq h_0$ from $v$, this subgraph is isomorphic to $G$.

Now, let us condition on event (*) {\it not} occurring for {\it any} $v$.  At the end of the proof of the lemma, we will upper bound the probability of event (*) occurring.
If event (*) does not occur, then the distribution of queries responses in the subtree of $v$ is the same whether $v$ has level $0$ or $1$.
So, when vertex $a$ is added to $S(q)$, it has probability $1/2$ of being $0$, conditioned on the level being $0$ or $1$, i.e., 
we toss an unbiased coin to determine the level of that vertex: ``heads" corresponds to level $0$ and ``tails" corresponds to level $1$.  The level may also be $>1$, but including that possibility only increases the number of queries.

The number of queries $q$ is at least equal to $Q$ times the number of ``tails" that have occurred.
Remark:
``heads" also implies that there were at least $Q$ queries of the subtree of some vertex, but if we include those queries due to ``heads" in the total number, we must be careful to avoid overcounting
 as those $Q$ queries of the subtree of some vertex $v_i$ will also give some number of queries (up to $Q-1$) of descendants of $v_i$.
So, for simplicity, we will use the lower bound that the number of queries is at least $Q$ times the number of tails.

Starting from the root of $G'$, to reach $S_{leaf}$ in $G'$ with at most $q$ queries, we must have one of these two possibilities: {\bf (1)} subtree $S(q)$ contains some vertex at level $0$ with distance $\geq H-h_0$ from the root; 
or {\bf (2)} we reach $S_{leaf}$ in $G'$ in fewer than $Q$ queries starting with some vertex $v$ of level $0$ and distance $\leq H-h_0$ from the root, i.e., we reach $S_{leaf}$ in $G'$ in the subtree of such a vertex $v$ with fewer than $Q$ queries in that subtree.

Conditioned on event (*) not occurring,
the probability of {\bf 1} occurring in at most $(M-(H-h_0))Q$ queries s bounded by the probability of having at least $H-h_0$ heads out of $M$ coin tosses and so is bounded by
$\Pbias(H-h_0,M)$.
If event {\bf 2} happens, then event (*) happens.
By a union bound, since there are only $Q'$ vertices that we query, event (*) happens with probability at most $Q' P$. 
So, by a union bound, the probability of reaching $S_{leaf}$ in $G'$ with at most $QM$ queries is bounded by
$\Pbias(H-h_0,M)+Q'P$.

Remark: likely the union bound in the previous paragraph on the probability of event (*) could be tightened.
If we query a vertex at distance $H-h_0$ from the root then indeed the probability of reaching $S_{leaf}$ is bounded by $P$ but for vertices of lower distance from the root the probability is less as would follow from a better inductive assumption.  We will not need this tightening so we omit it.

Having proved the first claim, the proof of the second claim is almost identical.
We use two new ideas.
The set $\Delta$ is at distance $\glen/2-O(1)$ from the start vertex of $D$.  So, to reach $\Delta$ one must at some time query some vertex $v$ which is at level $0$ and at distance $\lfloor \glen/4 \rfloor $ from the start vertex of $D$ such that in the subtree of $v$ one then queries a vertex in $\Delta$.

This introduction of vertex $v$ is the first new idea: by choosing such a vertex $v$ which is far from the start vertex, we will be able to ignore the possibility that the algorithm gets extra information about which vertex is the start vertex in response to queries by considering
only vertices near $v$.
Indeed, consider queries in the subtree of $v$.
Consider the set of vertices of $D$ including all vertices of level $\geq 1$ and all vertices of level $0$ which are
distance $\glen/4-O(1)$ so that this set does not intersect $\Delta$ and does not contain the start vertex.
Let $D'$ be the subgraph of $D$ induced by this set of vertices.
In the modified query model one cannot distinguish $D'$ from its universal cover $\tilde D'$, which is a tree graph; this is the second new idea.
Take the root of this tree graph to have its image under the covering map be $v$.
Then, responses to queries on this tree graph are the same as almost the same in queries on the graph given by decorating $T_{n,\glen/4-O(1)}$ 
$h_0,h_1,\ldots,h_k$.  The only minor difference is that we may take the first query to be nonbacktracking, so that rather than decorating graph $T_{n,\glen/4-O(1)}$ we decorate an $n$-ary tree of depth $\glen/4-O(1)$, i.e., there are $n$ rather than $n+1$ new neighbors in response to the first query.
To reach $\Delta$, one must reach some vertex which is at level $0$ and at distance $\glen/4-O(1)$ from $v$.

Then, the rest of the proof is the same.
\end{proof}
\end{lemma}

\section{Choice of Height Sequence and Proof of Main Theorem}
\label{chs}
We now make specific choices of the height sequence to prove \cref{mainth}.

We pick $n=N^8$ in the construction of $C,D$.  The value of $n$ does not matter for the classical lower bounds that follow from \cref{inductivelemma}.  However, the choice of $n$ does affect the spectrum of the quantum Hamiltonian.
We pick $\glen=4(l+1) N+O(1)$.
Thus $n/\glen^2=\Theta(N^6/l^2)$.

We decorate with height sequence $h_1,\ldots,h_l$, choosing $l=\lfloor \log_2(N)\rfloor $.
Thus, the graph $D$ is $d$-inner regular, with $d=\Theta(N^9)$.
We pick $$h_k=N \cdot (l+1-k),$$
for that $h_{k-1}-h_k=N$ and $L/4-O(1)-h_1=N$.

We first quickly show items {\bf 1-3,5} of \cref{trq} before showing the harder result, item {\bf 4}.  Then, \cref{mainth} follows from \cref{blemma}.

Note that $n/\glen^2\gg\sqrt{d}$.  So,
from \cref{eigbounddec}, it follows that the 
difference between the ground state energy of $C$ and $D$ is $\Theta(N/L^2))$ and so item {\bf 1} of \cref{trq} is satisfied.

Again since $n/\glen^2\gg\sqrt{d}$, from the bound on the second largest eigenvalue of the adjacency matrix of $C$ in \cref{eigbounddec}, we satisfy the condition on the spectral gap of the Hamiltonian of $C$, item {\bf 2} of \cref{trq}.

Item {\bf 3} of \cref{trq} is also satisfied by the last result in \cref{eigbounddec}, since there are only $O(N)$ original vertices and the amplitude of the ground state wavefunction is the same on all of them.

Item {\bf 5} of \cref{trq} trivially follows.  The number of vertices in $G_0$ is $\leq \poly(N)$.  Decoration by an $m$-ary tree of height $h$ multiplies the number of vertices in the graph by $O(m^h)$.  All $h_k$ are $O(N \log(N))$, and the largest $m$ is $O(\poly(N))$, and there are only $O(\log(N))$ steps of decoration, so the total number of vertices is $\exp(O(N \log(N)^3))$. 

Remark: since the construction of an admissible path in \cref{distg} uses $N$ copies of the graph, the number of computational basis states needed is $\exp(O(N^2 \log(N^3))$.  This number can be reduced somewhat since we could choose smaller values of $h_k$, such as $h_k=N^\alpha \cdot (l+1-k)$ for some smaller $\alpha<1$; we omit this.

To prove item {\bf 4}, we use \cref{inductivelemma}.
We first use the first part of the lemma.
Consider a sequence of graphs $J_l,J_{l-1},J_{l-2},\ldots,J_1$, where $J_k$ is given by decorating
$T_{n_k,h_k}$ by height sequence $h_{k+1},h_{k+2},\ldots,h_l$, for $n_k=n \cdot 2^k$.

Clearly, for $J_l$, it is $(0,N-1)$-hard to reach $S_{leaf}$, simply because the tree has height $N$.
We apply 
\cref{inductivelemma} to use $(P_{k+1},Q_{k+1})$-hardness on $J_{k+1}$ to show $(P_{k},Q_{k})$-hardness on $J_{k}$.
For all $k$, we pick $M=(4/3)N$, so that
\be
\label{Qk}
Q_k=(N/3)^{l-k} N.
\ee

Then $\Pbias=\exp(-\Omega(N))$ and
\cref{precursion}
gives that $P_{k-1}\leq \exp(-\Omega(N))+N P_k$.
Hence
\begin{eqnarray}
P_k &\leq & \exp(-\Omega(N))+Q_k P_{k+1} \\ \nonumber
&\leq & \exp(-\Omega(N))+Q_k \exp(-\Omega(N))+Q_k Q_{k+1} P_{k+2} \\ \nonumber
& \leq & \ldots \\ \nonumber
&\leq & \exp(-\Omega(N)) (1+Q_k+Q_k Q_{k+1}+\ldots) \\ \nonumber
&\leq & \exp(-\Omega(N)) O(Q_k^l) \\ \nonumber
& \leq & \exp(-\Omega(N)) \exp(O(\log(N)^3)) \\ \nonumber
& \leq & \exp(-\Omega(N)).
\end{eqnarray}

It is $(P,Q)$-hard to reach $S_{leaf}$ in $J_1$ with
\be
P=\exp(-\Omega(N)),
\ee
and
\be
Q=\exp(\Theta(\log(N)^2)).
\ee

Then, use the second part of the lemma, with the same $M$, to show
\begin{lemma}
\label{hardnesslemma}
It is $(P,Q)$-hard to reach $\Delta$ in $D$ with
\be
P=\exp(-\Omega(N)),
\ee
and
\be
Q=\exp(\Theta(\log(N)^2)).
\ee
\end{lemma}

This completes the proof of \cref{mainth}.

We make one final remark on the Hamiltonian on these decorated graphs.  For the given parameters, the ground state wavefunction on $C$ has most of its probability (its $\ell_2$ norm) on the original vertices of $C$.  However, the $\ell_1$ norm is concentrated near the terminal vertices.  The distinction between $\ell_1$ and $\ell_2$ norm was used in \cite{obs} to ``pin" the worldline in the case of path integral Monte Carlo with open boundary conditions and was considered in \cite{Jarret_2016} as an obstruction for diffusion Monte Carlo methods.   
The large $\ell_2$ norm of the ground state wavefunction on $C$ can be regarded as arising from all the short cycles on $C$; if we replace $C$ with a (finite) tree with the same degree, then ground state energy on $C$ shifts and the $\ell_2$ norm of the ground state becomes concentrated instead near the terminal vertices.  Thus, one may say that the topological obstructions are related to the $\ell_1$ versus $\ell_2$ obstructions.  The idea of our construction here is to make it so that no classical algorithm can (once one is far from the start vertex) efficiently distinguish original vertices from other vertices, since it cannot detect the short cycles between the original vertices and so it is unable to determine that the $\ell_2$ norm should be larger.

\section{Linear Paths}
\label{linear}
We have considered a path $H_s$ given by an oracle and we have implemented some complicated $s$-dependent terms so that certain terms increase and then later decrease over the path.  One might be interested in the case of linear interpolation so that $H_s=(1-s) H_0+sH_1$ for some fixed and known $H_0$, such as a transverse field, with $H_1$ being a diagonal matrix given by an oracle.  It seems likely that the construction here could be adapted to show hardness in that case too, using some perturbative gadgets.  Since it is not too interesting a question, we only sketch how this might be done.

One might induce dynamics on an unknown graph (either $C$ or $D$) as follows.
Both graphs have the same vertex set but have different edge sets.  Define a new graph $E$ which has one vertex for each vertex in $C$ and one vertex for each edge in $C$.  Given a pair of vertices $v,w$ connected by an edge $e$ in $C$, the graph $E$ will have
an edge between the vertex corresponding to $v$ and the vertex corresponding to $e$, as well as an edge between the
vertex corresponding to $w$ and the vertex corresponding to $e$.  We add also a diagonal term to the Hamiltonian on graph $E$ which assigns some positive energy to all vertices which correspond to edges of $C$.  If all these diagonal terms are the same and are chosen appropriately, we have a perturbative gadget which gives us an effective Hamiltonian corresponding to the Hamiltonian of graph $C$, up to an overall multiplicative scalar.  On the other hand, we could increase the diagonal term on vertices which correspond to edges of $C$ which are not edges of $D$ to effectively induce the Hamiltonian on $D$.

Also, in our construction of a path in \cref{distg} we used the ability to turn on and off a term $t|0\rangle\langle s| + h.c.$ in the Hamiltonian.  Such a term it seems can be induced by some perturbative gadgets also.  First  consider a slightly more general case where $H_0=(1-s) H_0 + H^{diag}_s$ where $H_0$ is fixed, known term and $H^{diag}_s$ is a diagonal matrix depending on $s$ and given by an oracle.  Then, we can effectively induce a $t$ which depends on $s$ by using hopping between an intermediate state, i.e., $-(|0\rangle\langle {\rm int}| + |{\rm int}\rangle\langle s|)+h.c.$ where $|{\rm int}\rangle$ is some intermediate state.  Then adding an $s$-dependent diagonal term on $|{\rm int}\rangle$ can be used to turn on or off the effective hopping between $|0\rangle$ and $|s\rangle$.

To effectively induce this $s$-dependent diagonal term on $|{\rm int}\rangle$ we could use another trick: replace every basis state with some set of basis states of size ${\rm poly}(N)$ and adding hopping terms between every pair of basis states in each such set. 
Then, each such set defines a subspace.
 Further add some diagonal terms on some of the basis states in each set, so that, for example, one set might include $n_0$ states with energy $E_0$, some number $n_1\gg n_0$ of additional states with energy $E_1\gg E_0$, and so on.  Then, if $s$ is close to $1$ so that $(1-s)H_0$ is small, almost of the amplitude will be in the states with energy $0$ and we can treat that set of states as a single state with energy $E_0-(1-s) (n_0-1)$; for slightly smaller $s$, we will start to occupy higher energy states and for some choices of sequences $n,E$ we can treat that set approximately as  a single state with energy
$\approx E_1-(1-s) (n_0+n_1-1)$, and so on. 
 By adjusting the sequences of $n,E$ on each set, it seems that we can 
 effectively implement a problem with fairly complicated $s$-dependent diagonal terms.

\bibliographystyle{unsrturl}
\bibliography{dec-ref}
\end{document}